\documentclass[useAMS,usenatbib]{mn2e}
\usepackage{epsfig}
\usepackage{times}
\usepackage{amsmath}

\newif\ifAMStwofonts

\newcommand{\target}{PSR\,J1023+0038}
\newcommand{\Msun}{$M_{\odot}$}
\newcommand{\Rsun}{$\rm \,R_{\odot}$}
\newcommand{\Porb}{$P_{\rm orb}$}

\newcommand{\rma}{$r_{\rm m}$}
\newcommand{\rco}{$r_{\rm cor}$}
\newcommand{\rlc}{$r_{\rm lc}$}

\newcommand{\ergs}{\,$\rm erg\,s^{-1}$}

\newcommand{\sloanu}{$\it u'$}
\newcommand{\sloang}{$\it g'$} 
\newcommand{\sloanr}{$\it r'$}
\newcommand{\sloani}{$\it i'$}

\title[Optical observations of PSR\,J1023+0038] 
{The binary millisecond pulsar PSR\,J1023+0038 during its accretion state -- I. 
Optical variability 
}

\author[T.\,Shahbaz et al. ]
       {T.\,Shahbaz,$^{1,2}$\thanks{E-mail: tsh@iac.es}
		M.\,Linares,$^{1,2}$
		S.\,P.\,\,Nevado,$^{1,2}$ 
		P.\,Rodr\'\i guez-Gil,$^{1,2}$
		J.\,Casares,$^{1,2}$  \newauthor   
		V.\,S.\,Dhillon,$^{3,1}$
		T.\,R.\,Marsh,$^4$
		S.\,Littlefair,$^3$
		A.\,Leckngam,$^{5}$
        	S.\,Poshyachinda,$^{5}$\\  
$^{ \it 1}$Instituto de Astrof\'\i{}sica de Canarias (IAC), E-38200 La Laguna, Tenerife, Spain \\
$^{ \it 2}$Dept. Astrof\'\i{}sica Universidad de La Laguna (ULL), E-38206 La Laguna, Tenerife, Spain \\
$^{ \it 3}$Department of Physics and Astronomy, University of Sheffield, Sheffield, S3 7RH, UK  \\
$^{ \it 4}$Department of Physics, University of Warwick, Coventry CV4 7AL, England \\
$^{ \it 5}$National Astronomical Research Institute of Thailand, 191 Siriphanich Building, 
Huay Kaew Road, Chiang Mai 50200, Thailand
}

\pagerange{\pageref{firstpage}--\pageref{lastpage}}
\pubyear{2014}

\begin{document} 
\maketitle 
\begin{abstract} 
\noindent

We present time-resolved optical photometry of the binary millisecond `redback'
pulsar \target\ (=AY\, Sex) during its low-mass X-ray binary phase. The light
curves taken between 2014 January and April show an underlying sinusoidal
modulation due to the irradiated secondary star and accretion disc. We also
observe superimposed rapid flaring on time-scales as short as $\sim $20\,s with
amplitudes of $\sim 0.1-0.5$\,mag and  additional large flare events on
time-scales of  $\sim 5-60$\,min with amplitudes of $\sim$0.5-1.0\,mag. The
power density spectrum of the optical flare light curves is dominated by a 
red-noise component, typical of aperiodic activity in
X-ray binaries. Simultaneous X-ray and UV observations by the Swift satellite
reveal strong correlations that are consistent with X-ray reprocessing of the UV
light, most likely in the outer regions of the accretion disc.
On some nights we also observe sharp-edged, rectangular, flat-bottomed dips 
randomly distributed in orbital phase, with a median duration of $\sim$250\,s
and a median ingress/egress time of $\sim 20$\,s. These rectangular dips are
similar to the mode-switching behaviour between disc `active' and  `passive' 
luminosity states,  observed in the X-ray light curves of other redback
millisecond pulsars. This is the first time that the optical analogue of the
X-ray mode-switching has been observed. 
The properties of the passive  and active
state light curves can be explained in  terms of 
clumpy accretion from a trapped inner
accretion disc near the corotation radius,  resulting in rectangular, 
flat-bottomed optical and X-ray light curves.

\end{abstract}

\begin{table*}
\caption{Log of optical observations. The mean observed magnitude is given in
the final column. The difference between the exposure time and time resolution
is due the readout time of the detector.}
\begin{center}
\begin{tabular}{llccccccc}\hline 
Telescope & \textsc{ut} Date &  Start HJD-2450000 & Duration (h)  & Exp time (s)  & Time res. (s)  & Filter  & Seeing \arcsec & Mag \\
\hline
TNT     & 2014/01/29 & 6687.18760 & 1.71  & 0.97 & 0.97 & r' & 1.2 & 16.63 \\
TNT     & 2014/01/30 & 6688.28886 & 4.51  & 0.97 & 0.97 & r' & 1.3 & 16.16 \\
IAC80   & 2014/02/02 & 6691.66953 & 2.53  & 15 & 30  & r' & 1.4 & 16.31 \\
IAC80   & 2014/02/03 & 6692.55923 & 5.02  & 15 & 30  & r' & 1.5 & 16.38 \\
IAC80   & 2014/02/04 & 6693.54567 & 5.24  & 15 & 30  & r' & 1.4 & 16.61 \\
IAC80   & 2014/02/06 & 6695.58626 & 4.15  & 20 & 47  & r' & 1.9 & 16.79 \\ 
IAC80   & 2014/03/01 & 6718.39729 & 7.43  & 15 & 30  & r' & 1.7 & 16.77 \\
IAC80   & 2014/03/06 & 6723.39929 & 6.06  & 30 & 149 & r' & 1.7 & 16.80 \\
IAC80   & 2014/03/08 & 6725.39641 & 5.37  & 30 & 45  & r' & 1.9 & 16.56 \\
WHT     & 2014/03/16 & 6733.35428 & 0.88  & 0.28 & 0.31 & r' & 1.3 & 16.39 \\
WHT     & 2014/03/16 & 6733.35428 & 0.88  & 0.28 & 0.31 & g' & 1.5 & 17.40 \\
WHT     & 2014/03/16 & 6733.35428 & 0.88  & 1.53 & 1.55 & u' & 1.7 & 17.12 \\
IAC80   & 2014/03/16 & 6733.36879 & 6.80  & 30 & 45  & R  & 1.6 & 16.21 \\
IAC80   & 2014/03/17 & 6734.38586 & 6.34  & 30 & 45  & R  & 1.8 & 16.66 \\
TNT     & 2014/03/25 & 6742.06245 & 6.19  & 3.37 & 3.37 & r' & 1.0 & 16.60 \\
IAC80   & 2014/03/25 & 6742.35805 & 4.15  & 30 & 45  & R  & 2.7 & 16.52 \\
IAC80   & 2014/03/26 & 6743.34121 & 6.60  & 30 & 45  & R  & 1.8 & 16.34 \\
TNT     & 2014/03/28 & 6745.10920 & 4.98  & 0.97 & 0.97 & r' & 1.1 & 16.26 \\       
IAC80   & 2014/03/29 & 6746.35337 & 6.37  & 30 & 45  & R  & 1.6 & 16.43 \\
INT     & 2014/04/08 & 6756.35347 & 5.39  & 10 & 39  & r' & 2.7 & 16.61 \\ 
IAC80   & 2014/04/24 & 6772.37623 & 4.08  & 30 & 45  & r' & 1.3 & 16.43 \\
WHT     & 2014/04/29 & 6777.35771 & 2.17  & 0.28 & 0.31 & r' & 1.0 & 16.40 \\
WHT     & 2014/04/29 & 6777.35771 & 2.17  & 0.28 & 0.31 & g' & 1.0 & 17.15 \\
WHT     & 2014/04/29 & 6777.35771 & 2.17  & 1.53 & 1.55 & u' & 1.1 & 17.40 \\
\hline
\end{tabular}
\end{center}
\label{log}
\end{table*}

\begin{figure*}
\centering
\includegraphics[angle=0,height=23.cm]{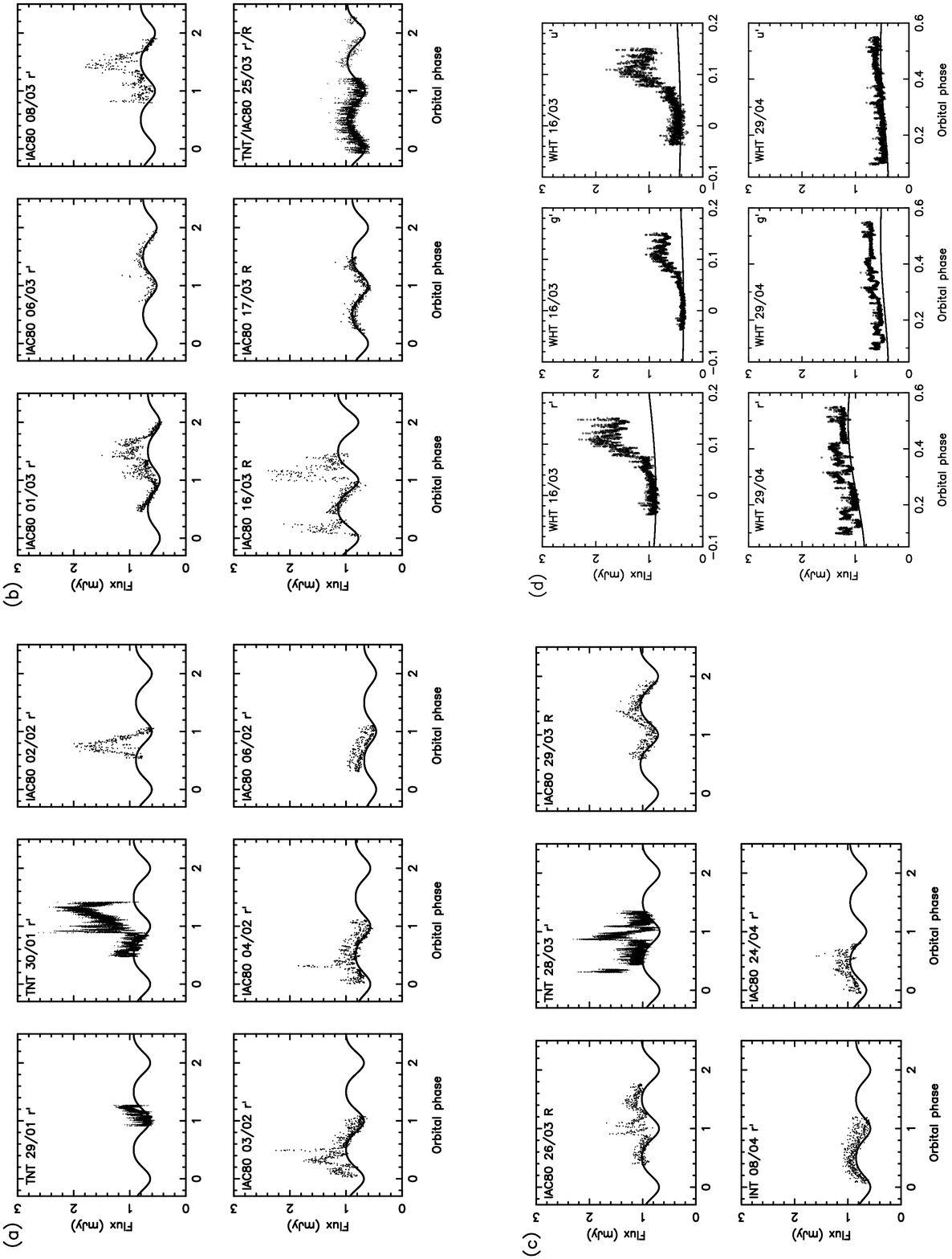}
\caption{Observed optical light curves of \target\ taken with the IAC80, INT and TNT  (a-c) and 
the WHT (d). The light curves have been phase-folded using the ephemeris given in Section\,\ref{lcurves}; phase 0.0 is defined as inferior conjunction of the secondary star. The day/month and filter are also given in each panel. The solid line is a sinusoidal modulation  (the secondary star's ellipsoidal modulation plus heated star and accretion disc)  scaled to fit the lower envelope (see Section\,\ref{lcurves}).
}
\label{lcurve}
\end{figure*}

\begin{figure*}
\centering
\includegraphics[angle=0,height=23.cm]{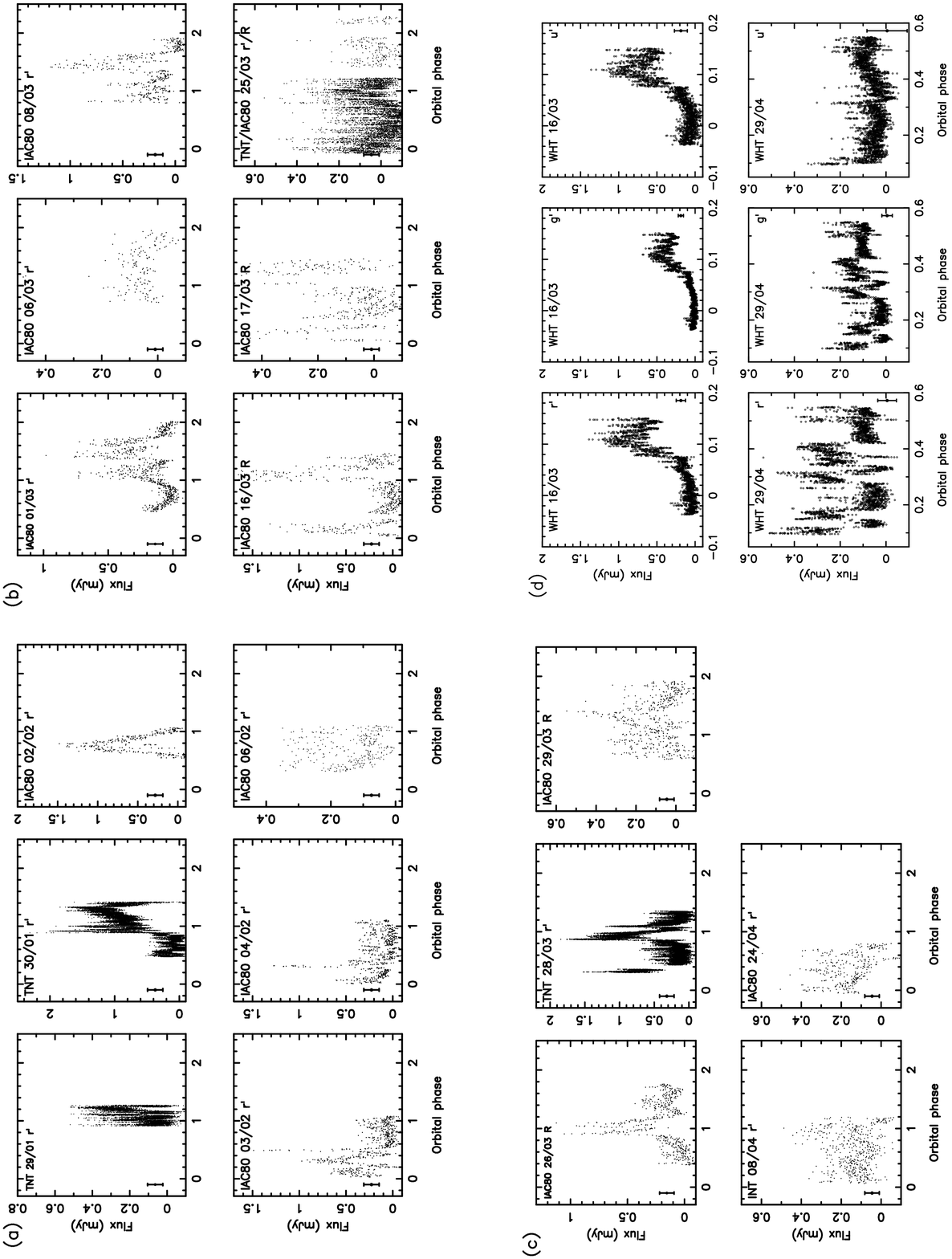}
\caption{The de-reddened and de-trended optical light curves of \target, 
obtained by subtracting a fit to the lower-envelope of the light curves 
shown in Fig.\,\ref{lcurve}. (a-c) and (d) show the data taken with the IAC80, INT, TNT  and 
the WHT respectively. The solid circle marks the typical uncertainty  in the data. The date day/month and filter are also given in each panel. }
\label{detrend}
\end{figure*}

\section{INTRODUCTION}

The Galactic source PSR\,J1023+0038 (=AY\,Sex) was discovered by \citet{Bond02}.
It was initially classified as a magnetic cataclysmic variable, but subsequent
radio  studies  revealed  the true nature of the compact object was revealed,
with the discovery of  1.69\,ms  radio pulsations, establishing without a doubt
that the compact  object was a rotation-powered millisecond pulsar
\citep{Archibald09}.

Millisecond pulsars are formed in X-ray binaries by accreting gas from a
low-mass  companion star \citep{Alpar82}.  The discovery of accreting
millisecond X-ray pulsars \citep{Wijnands98,PW12} represented a  confirmation of
this `recycling' scenario  for the long-suspected evolutionary connection
between low-mass X-ray binaries and rotation-powered recycled millisecond 
pulsars. The accreted matter spins up the neutron star to millisecond periods,
but how and when the accretion process turns off remains unknown  
\citep[see][]{Tauris12}. \target\ has been heralded as the `missing link'
between radio millisecond pulsars and low-mass X-ray binaries and has given new
insights into this evolutionary process.

In 2001 strong evidence for the presence of an accretion disc in \target\ was
provided by optical observations of double-peaked emission lines, accompanied by
short-time-scale  flickering and a blue optical spectrum
\citep{Bond02,Szkody03}. From 2002 onwards optical/X-ray observations did not
show the presence of an accretion disc
\citep{Archibald09,Archibald10,Bogdanov11}.  Therefore, in 2001, \target\ was an
accreting neutron star in a low-mass X-ray binary. Optical spectra taken in
2003/2004 showed only the absorption lines from the secondary star with no
detectable emission lines and the radial velocity of the late-type G5 companion
star was found to be modulated on the 4.754\,h  orbital period (\Porb;
\citealt{Thorstensen05}).  Optical light curves taken in 2004 revealed a single
humped modulation which can be explained in terms of an X-ray heated optical
light curve \citep{Thorstensen05}. Between 2001 and 2007 radio pulsations were
found \citep{Archibald09} and until 2013 it has been consistently observed as an
eclipsing radio millisecond pulsar, where the radio eclipses are attributed to
ionized material in the companion's magnetosphere  \citep{Archibald13}. Given
the $\sim 0.2$\Msun\ secondary star, \target\ is known as a `redback' system 
\citep{Roberts11}, with a lower X-ray luminosity in the `pulsar state' due to  a
pulsar-wind-driven shock near the inner Lagrangian point \citep{Bogdanov11}.

Since 2013 June, \target\ switched off as a radio pulsar
\citep{Stappers13,Patruno14} and when solar constraints were overcome, optical
spectroscopy in mid-October revealed strong double-peaked H$\alpha$ emission
indicating that an accretion disc had re-formed \citep{Halpern13,Linares14a}. 
Moreover, observations showed that the X-ray emission had increased by a factor
of at least 20 compared to previous quiescent values \citep{Kong13} and the UV
emission had brightened by 4 magnitudes \citep{Patruno14}, with no detection of
radio pulsations. All these factors provide compelling evidence for the presence
of a recently formed accretion disc,  indicating that \target\ has switched back
from a radio millisecond pulsar to a low-mass X-ray binary.  The X-ray light
curves during the accretion phase show rectangular flat-bottomed dips
\citep{Tendulkar14} that are similar to the mode-switching behaviour between
disc `active' and  `passive'  luminosity states observed in other redback binary
pulsars \citep{Linares14c}. \citet{Coti14}  determined the near-infrared  to the
X-ray  spectral energy distribution (SED)  and find that it is well represented
by a model consisting of an irradiated companion, an accretion disc and 
intra-binary shock emission. The authors suggest a scenario in terms of an
engulfed radio pulsar which is still active, but is undetectable at radio
wavelengths  due to a large amount of ionized material surrounding the compact
object, material which is however not enough to totally quench the radio
pulsar. 

In this paper we report on  long-term and  high time-resolution optical
photometry obtained during the accretion disc/radio-quiet phase of \target\  in
2014. In a following paper we will present the result of the spectroscopic
campaign \citep{Linares15}.

\section{OPTICAL OBSERVATIONS AND DATA REDUCTION}

A total of 24  time-resolved optical light curves  of \target\ were obtained
using four telescopes; the 0.8-m IAC80 telescope (Tenerife, Spain),  the 2.4-m
Thai National Telescope (TNT; Thai National Observatory, Thailand), the 2.5-m
Isaac Newton Telescope (INT; La Palma, Spain) and  the 4.2-m William Herschel
Telescope (WHT; La Palma, Spain), during the period 2014 January to April.
Conventional CCD optical light curves of \target\ were obtained with the
IAC80+CAMELOT and the INT+WFC with plate scales of 0.30 and 0.33\,\arcsec/pixel,
respectively. The observing conditions were variable with seeing ranging from 1
to 2\,\arcsec. The exposure times used ranged from 10 to 60\,s with the \sloanr\
or $R$ filter depending on  availability.  For the WFC data the CCD was windowed
to decrease the read-out time. Bias images and flat-fields were also taken. We
used \textsc{iraf}\footnote{\textsc{iraf} is distributed by the National
Optical  Astronomy Observatory, which is operated by the Association of
Universities for Research in Astronomy, Inc., under cooperative agreement with
the National Science Foundation. http://iraf.noao.edu/} for our data reduction,
which included bias subtraction images and flat-fielding. The \textsc{vaophot} 
software \citep{Deeg01} was used to obtain differential light curves for
\target\ and several comparison stars by extracting the counts using variable
aperture photometry which scaled with the seeing. High time-resolution light
curves were obtained at the TNT with  ULTRASPEC \citep{Dhillon14} and at the WHT
with  ULTRACAM \citep{Dhillon07}. ULTRASPEC has an electron multiplier
frame-transfer CCD with a pixel scale of 0.45\,\arcsec/pixel and allows for
single-band, high-speed imaging. ULTRACAM  has three  frame-transfer CCDs with a
pixel scale of 0.3\,\arcsec/pixel, allowing one to obtain simultaneous blue,
green and red-band high-speed images. Due to the frame-transfer architecture of
the CCDs, the dead-time is essentially zero with ULTRACAM and ULTRASPEC.  With
ULTRASPEC we obtained \sloanr-band images with  exposure times of 0.96 and
3.35\,s, with a typical photometric accuracy of 4.9 and 6.7 per cent,
respectively. With ULTRACAM we obtained simultaneous \sloanu, \sloang\ and
\sloanr-band images with an exposure time of 0.28\,s in \sloang\ and \sloanr\,
and 1.53\,s in \sloanu, with a typical photometric accuracy of 6.0, 7.0 and 5.1
per cent respectively. The ULTRACAM pipeline reduction software was used to
debias and flat-field the data. The same pipeline was also used to obtain light
curves for \target\ and several comparison stars by extracting the counts using
aperture photometry with variable apertures that scaled with the seeing.

For all of the data, calibrated magnitudes of the target with respect to a
bright local standard SDSS\,J102343.30+003819.1  (\sloanu=16.810,
\sloang=15.676, \sloanr=14.773 and \sloani=14.952) were then determined.  We
checked that the local  standard was not variable by comparing it to other
bright stars in the field.  For the data taken with the $R$ filter, we used the
SDSS transformations
\footnote{http://www.sdss.org/dr5/algorithms/sdssUBVRITransform.html} to
transform the \sloanr\ magnitude of the local standard to $R$. We checked the
transformation by using the IAC80 $R$ and WHT \sloanr\ light curves taken on
March 26, which overlapped in time. We binned the light curves to the same time
resolution and found both light curves to match with an offset of less  than 2
per cent. As a check of the photometry and systematics in the reduction
procedure, we also extracted light curves of a  comparison star of similar 
brightness to \target, which showed no significant variability above what is
expected from photon statistics.

\begin{figure}
\centering
\includegraphics[angle=-90,width=8.cm]{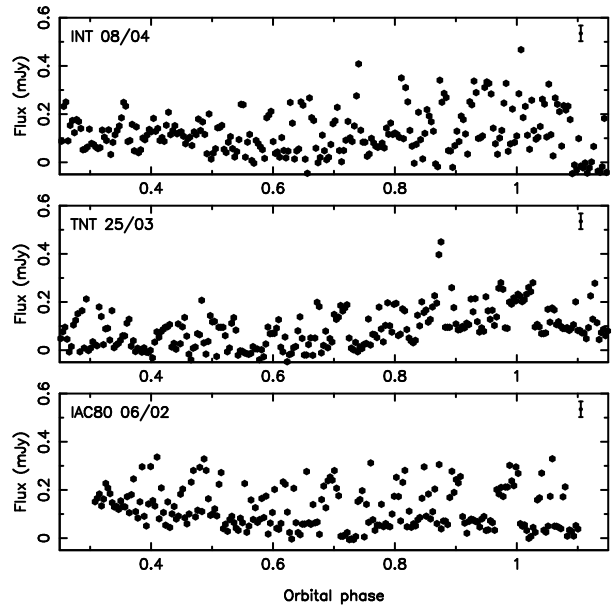}
\caption{The February 6 (bottom), March 25 (middle) and April 8 (top) \sloanr-band de-reddened and de-trended light curves. The March 25 light curve taken with  TNT+ULTRASPEC  has been re-binned to the same time-resolution as the IAC80 and INT light curves. The variability represents genuine variations of the target as indicated by the error bar plotted in the top right hand corner of each panel.} 
\label{short_flares}
\end{figure}

\section{OPTICAL LIGHT CURVES}
\label{lcurves}

We determined the colour excess by using the empirical relation between 
the colour excess and the total column density of hydrogen $\rm N_{\rm 
H}$ \citep{PS95}. Using the observed value for $\rm N_{\rm H}=5.2\times 
10^{20}\,cm^{-2}$ \citep{Coti14} and the ratio $A_{\rm V}/E(B-V)$=3.1 
\citep{Cardelli89} we obtain a colour excess of E(B-V)=0.073.
Fig.\,\ref{lcurve} shows the de-reddened flux IAC80, INT, TNT and WHT 
light curves where we have used the colour excess determined above 
and then converted the magnitudes to flux density using the
appropriate zeropoint for each band. 
We have also phase folded the data using the
ephemeris \Porb=0.1980963569\,d and $\rm T_{\it 0}$=2456642.6335
\,HJD\citep{Archibald09,Linares15}, where phase 0.0 is defined as inferior
conjunction of the secondary star.

\subsection{Flaring activity}
\label{flaring}

The light curves show dramatic variations from night to night superimposed on an
underlying sinusoidal modulation, which is due to a combination of the secondary
star's ellipsoidal modulation and it's heated inner face \citep{Thorstensen05}. 
On some nights there is additional  rapid flaring (February 6; March 25; April
8) and on other nights there are further  additional large flare events (January
1, February 4, March 8). In order to isolate the flaring activity (by removing
the shape of the secondary star's modulation) we subtract a scaled version of
the underlying sinusoidal modulation, defined by fitting the lower-envelope of
the light curve taken on March 17 with a sine-wave. The March 17 light curve was
chosen because it clearly showed the sinusoidal modulation with little flaring
behaviour. Fig.\,\ref{detrend}  shows the de-reddened and de-trended IAC80, INT,
TNT and WHT flare light curves. 

\begin{figure*}
\centering
\includegraphics[angle=0,width=16.cm]{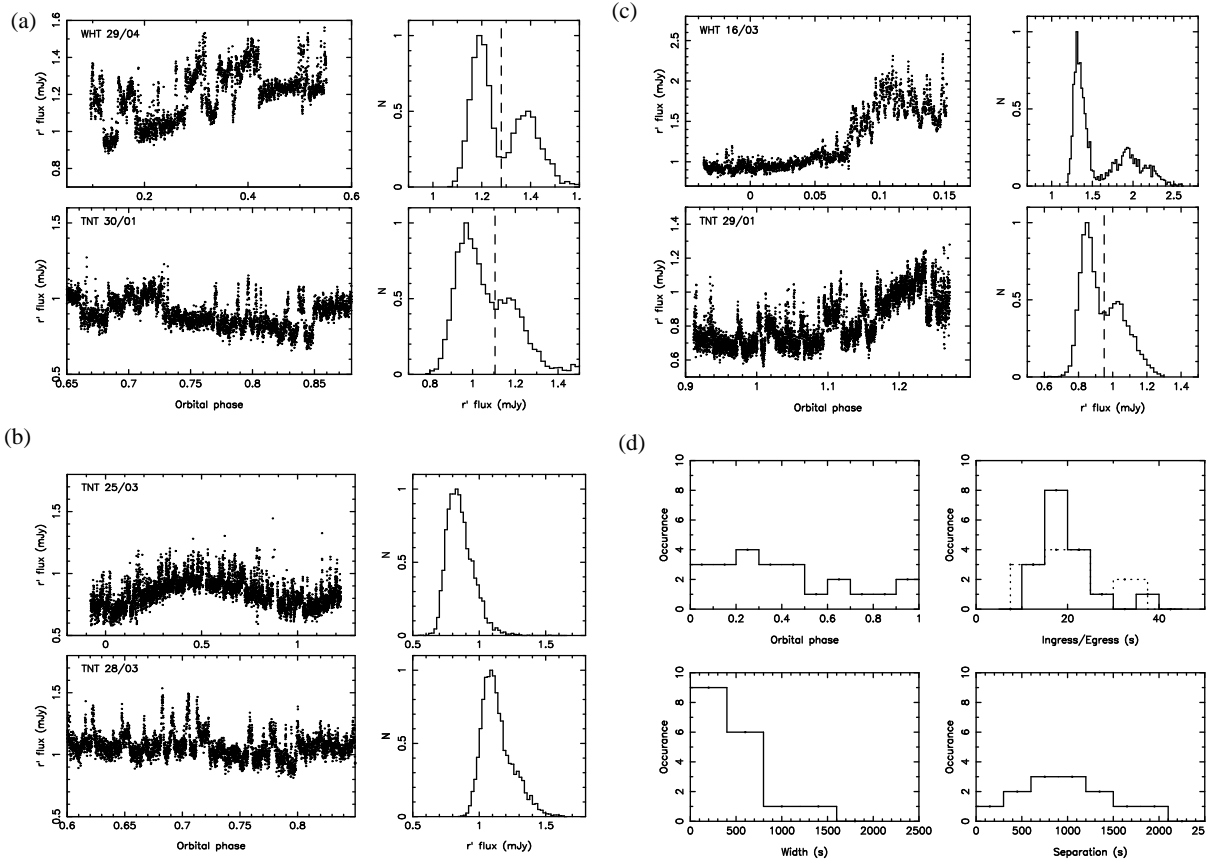}
\caption{(a-c) The de-reddened \sloanr-band light curves of \target, taken with the TNT+ULTRASPEC and the WHT+ULTRACAM.  The right panel in each plot shows a histogram of the flux values after correcting for the secondary star's trend. The mode-switching active  and passive state behaviour is only observed in the January 29, 30 and  April 29  light curves. The vertical dashed lines shown marks the cut-off flux value that defines the active flare and  active passive respectively. (d) Histograms of the observed rectangular mid-dip phase, ingress/egress times, duration and the separation between consecutive dips. }
\label{subsec}
\end{figure*}

Fig.\,\ref{short_flares} shows the flares on February 6, March 25 and April 8.
The variability represents genuine variations of the target. Noticeable from the
IAC80 and INT data are numerous rapid flare events that are not resolved.  The
TNT March 25 data were taken with ULTRASPEC and so have a much higher
time-resolution than the IAC80 and INT data (see Table\,\ref{log}) and so the
flares are better resolved. In order to directly compare the light curves, the
March 25 data were re-binned to the same time-resolution as the other data. The
subsequent light curves obtained are very similar and hence we can conclude that
the IAC80 and INT data do not resolve the rapid flares. The TNT data show rapid
flare events on time-scales of $\sim$minutes with amplitudes of $\sim
0.1-0.5$\,mag (see Fig.\,\ref{subsec}). Large flare events on time-scales of
$\sim5-60$\,min with amplitudes of $\sim 0.5-1.0$\,mag are also seen (e.g.
February 2, March 1) superimposed on the rapid flare events. In
Table\,\ref{PLindex} we give the fractional root-mean-square (RMS) of the
variability,  determined after binning the light curves to the same
time-resolution of 60\,s.

\begin{figure*}
\centering
\includegraphics[angle=0,height=8.0cm]{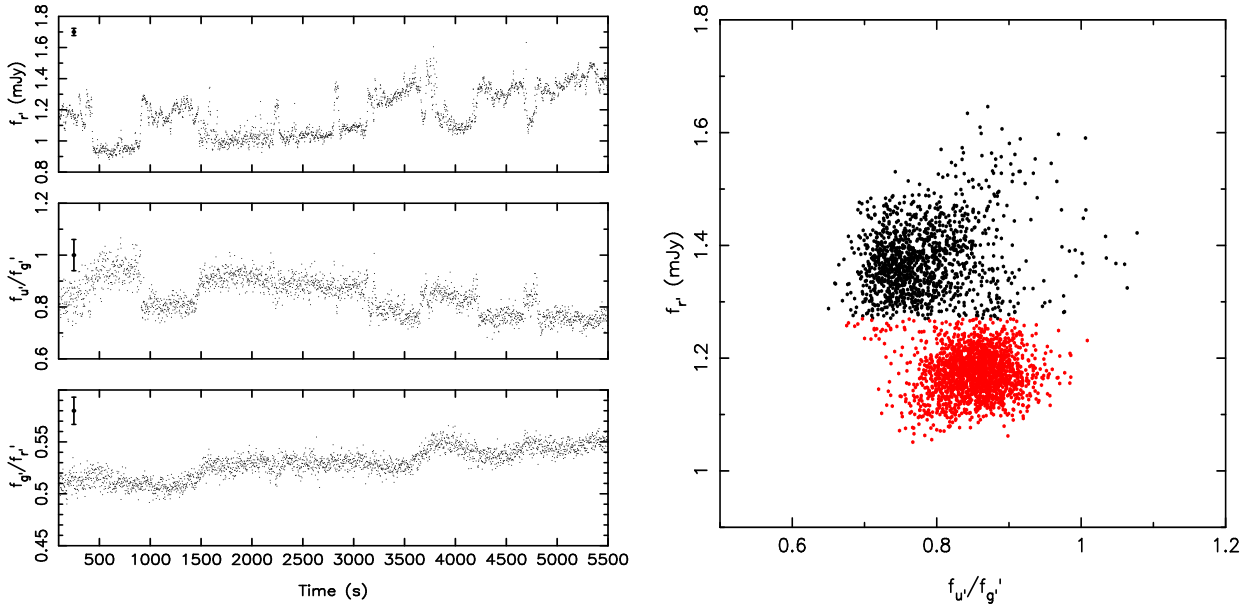}
\caption{Left: A close-up of the flat-bottomed dip features observed in the April 29 WHT light curve. We show the de-reddened \sloanr and the flux ratio $f_{u'}/f_{g'}$ and $f_{g'}/f_{r'}$ light curves. Notice the increase in the blue colour during the dips. The error bar marks the typical uncertainty  in the data. Right: Colour-magnitude diagram for the April 29 WHT data. We show the de-reddened \sloanr and the flux ratio $f_{u'}/f_{g'}$ and $f_{g'}/f_{r'}$ data. The solid black and red points show the active  and passive state data respectively. Note the increase in colour during the dips, i.e. they become more blue.}
\label{zoom}
\end{figure*}

\subsection{Rectangular dips}
\label{dips}

Fig.\,\ref{subsec} shows the observed \sloanr-band light curves and flux
histograms taken at the highest time-resolution; 0.31 s for the WHT data and
0.97 or 3.37\,s for the TNT data. These light curves, taken on January 29, 30
and April 29, show rectangular, flat-bottomed dip features which are similar to
the mode-switching behaviour (passive state and  active state) that have been
observed in the X-ray light curves of \target\ \citep{Tendulkar14} and other
X-ray binary pulsar \citep[see][and references therein]{Linares14c}. 
The
histogram of the flux values (using the flare light curves described in
Section\,\ref{lcurves} with the mean de-reddened flux added to preserve the flux
values)  clearly show a bimodal distribution between the passive  and active
-states (see Fig.\,\ref{subsec}).

In total we identify 16 rectangular dips in the January 29, 30 and April 29 
light curves. In order to estimate the properties of these dips, we fit each 
feature with a rectangular function, allowing us to determine the ingress,
egress, duration and centre of the dip segment. We find that the dips  are
generally symmetrical with ingress and egress times in the range 12 to 35\,s
(median of $\sim 20$\,s), a dip duration in the range 80 to 1300\,s (median of
$\sim 250$\,s) and a dip separation in the range 200 to 1900\,s (median of $\sim
900$\,s). We also compute histograms of the observed dip phase, the duration of
the dips and the separation between consecutive dips (see Fig.\ref{subsec}),
however, the low number of dips observed implies that we cannot comment on the
statistical properties of the dips with confidence.  In Fig.\,\ref{zoom} we show
a zoom of a typical rectangular dip feature observed on  April 29 with
ULTRACAM.  The feature is flat-bottomed with an increase in the  $f_{u'}/f_{g'}$
and  $f_{g'}/f_{r'}$ colour ratios during the dip, i.e. the dips become more
blue. We also show the colour-magnitude diagram of the whole light curve. The
increase in colour ratio during the dips is clearly seen.

\begin{figure*}
\centering
\includegraphics[angle=0,height=8.cm]{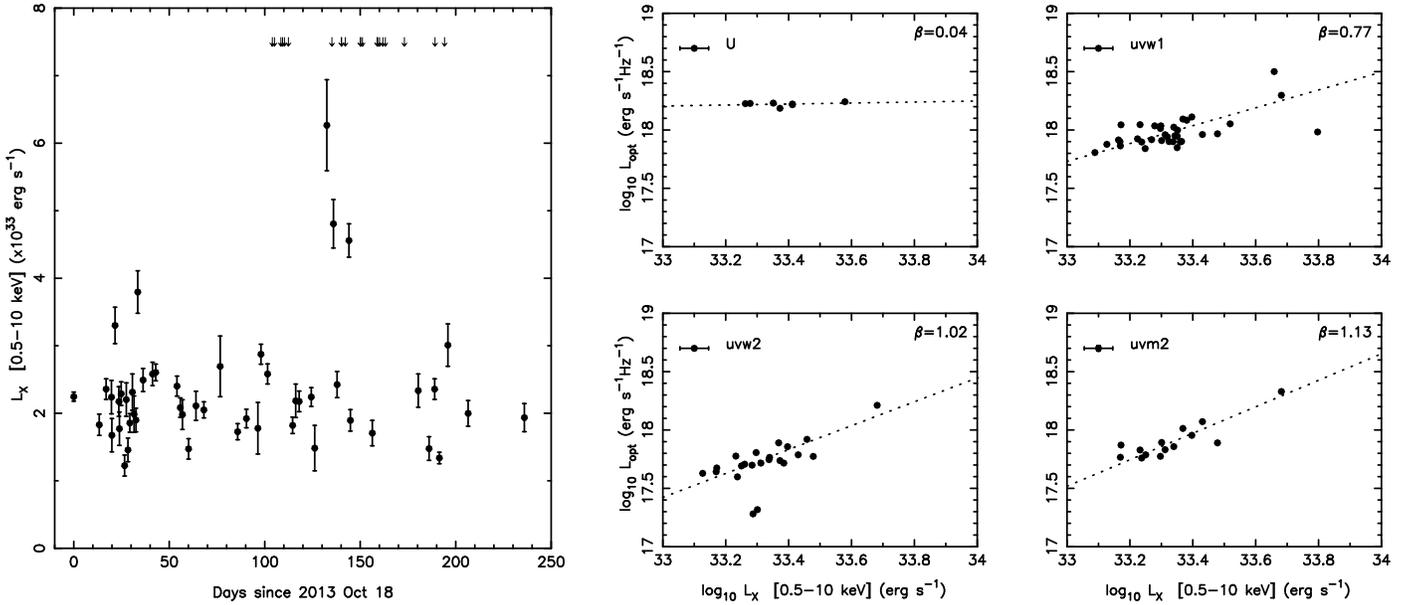}
\caption{Swift XRT and UVOT light curves of \target\ between 2013 Oct 18 and 2014 June 11. Left: Swift XRT 0.5--10\,keV light curve. The arrows mark the time of our IAC80, INT, TNT and WHT optical data. Right: Swift simultaneous UVOT (\textit{U}, \textit{uvw}1, \textit{uvw}2 and \textit{uvm}2) and XRT 0.5--10\,keV observations. 
 The error bar plotted in the top left hand corner of each panel. Tthe uncertainty in the UVOT data is smaller than the symbol size. A power-law fit to the data of the form 
$L_{\rm \nu; UV} \propto L_{\rm X}^{\beta}$  is shown. }
\label{lxlopt}
\end{figure*}

\section{X-RAY/ULTRAVIOLET  LIGHT CURVES}

We analyzed all {\it Swift} observations of PSR~J1023+0038 in the ongoing
accretion phase, taken between 2013 October 18 and 2014 June 11, including data
from both the X-ray \citep[XRT,][]{Burrows05} and UV/optical
\citep[UVOT,][]{Roming05} telescopes. This amounts to 52 observations, an
accumulated XRT exposure of 94\,ks and a total of 84 UVOT images acquired in
four different filters. Smaller sets of {\it Swift} observations of \target\
have been analyzed elsewhere \citep{Patruno14,Takata14,Linares14c,Coti14}. We
used the quasi-simultaneous XRT/UVOT average flux per observation (which lasts
typically between 1 and 3\,ks), as we are interested in the long-term
(weeks-months) X-ray-UV correlation. Because the UVOT filter configuration and
total XRT exposure vary, each observation had UVOT data collected in 1 to 3
different filters.

All XRT observations were taken in photon counting (PC) mode. For the three
brightest observations (on 2014 February 27,  March 3 and 11, during which the
count rate reaches $\sim$0.6~cts~s$^{-1}$), we used a 20-pixel radius extraction
region and excluded the innermost pixels (within a 2-pixel radius) to avoid
pile-up. The count rate during the remaining observations is $<$0.5~cts~s$^{-1}$
(below the PC pile-up threshold) and we used a 30\,arcsec radius circular
extraction region. We subtracted the background using a nearby 60\,arcsec radius
source-free region, and grouped the source spectra to a minimum of 15 counts per
channel. We created ancilliary response files for each observation, applying the
vignetting correction and using the corresponding exposure maps. After
incorporating the latest XRT response matrices, we fitted each spectrum in the
0.5--10~keV band using \textsc{xspec} \citep[v. 12.7.1,][except two spectra
which had less than 45 source counts and were not fitted]{Arnaud96}. We used a
power--law model corrected for absorption \citep[with
\textsc{tbabs},][]{Wilms00}, fixing the column density at the value measured by
\citet{Coti14} from the average XRT spectrum ($N_{\rm
H}$=5.2$\times$10$^{20}$~cm$^{-2}$). We calculated the 0.5--10\,keV X-ray
luminosity from the unabsorbed flux in the same band using the distance of
1.37\,kpc \citep{Deller12}.

After checking that the aspect correction was applied, we summed all extensions
within each UVOT image mode file (using the {\it Swift} ftool
\textsc{uvotimsum}, v. 1.6) to obtain one image per filter and observation.
Because our ULTRACAM and ULTRASPEC data provide much better measurements of the
fast optical variability of \target\ (see Section\,\ref{lcurves}), we do not
include in this analysis UVOT high-time resolution (event mode) data
\citep{Coti14}. We then performed aperture photometry on each summed image using
the standard 5\,arcsec aperture radius, and a nearby source-free background
region of 12\,arcsec radius.

In Fig.\,\ref{lxlopt} we show the \textit{Swift} XRT 0.5--10\,keV unabsorbed
long-term X-ray light curve. The X-ray light curve shows considerable activity
with an average luminosity of $L_{\rm X}\sim2\times10^{33}$\,\ergs. To search
for possible X-ray/UV correlations in Fig.\,\ref{lxlopt} we determine  the
simultaneous unabsorbed X-ray 0.5--10\,keV and de-reddened $U$, \textit{uvw}1, 
\textit{uvm}2  and \textit{uvw}2-band observations. The UV data were de-reddened
using the colour excess derived in Section\,\ref{lcurves}. The X-ray and
\textit{uvw}1,  \textit{uvm}2  and \textit{uvw}2-band data are clearly
correlated, with a  significance of $>$99 per cent. 
{However, given that there are only 7 points in the X-ray-\textit{U} 
correlation, the correlation is only significant at the 50 per cent level.
We fit the X-ray/UV data
with a power--law of the form $L_{\rm \nu; UV} \propto L_{\rm X}^{\beta}$ and
determine the power--law index (see Table\,\ref{index:lxlopt}). 

\begin{table}
\caption{Properties of the simultaneous X-ray/UV observations of \target. $N$ is the number of simultaneous X-ray/UV observations, $R$ is the correlation coefficient and $\beta$ is the power-law index of the form
$L_{\rm \nu; UV} \propto L_{\rm X}^{\beta}$.}
\begin{center}
\begin{tabular}{lccc}\hline 
Band              &  $N$   &  $R$       & $\beta$ \\
\hline 
\textit{U}         &  7   & 0.26     & 0.04(4) \\
\textit{uvw}1   & 35  & 0.62    & 0.77(2) \\
\textit{uvm}2   & 14  & 0.89   & 1.13(4) \\
\textit{uvw}2   & 22  & 0.87    & 1.02(3) \\
\hline       
\end{tabular}
\end{center}
\label{index:lxlopt}
\end{table}

\begin{table}
\caption{The fractional RMS of the $r'$-band flare light curves  and power-law (PL)  fits to the power density spectra. 
The light curves where active  and passive state are seen are not included. The mean 
de-reddened flux values are given and the fractional 
RMS values have been determined after binning the light curves  to the same time-resolution of 60\,s.}
\begin{center}
\begin{tabular}{llccc}\hline 
Telescope       & \textsc{ut} Date    &  Flux (mJy)  & RMS  (\%)     & PL index   \\
\hline
TNT      & 2014/01/29 &  0.94 &   9.5 & -1.43$\pm$  0.04 \\
TNT      & 2014/01/30 &  1.53 &  30.0 & -1.54$\pm$  0.04 \\
IAC80    & 2014/02/02 &  1.29 &  30.1 & -1.53$\pm$  0.07 \\
IAC80    & 2014/02/03 &  1.13 &  21.7 & -1.40$\pm$  0.04 \\
IAC80    & 2014/02/04 &  0.91 &  21.1 & -1.39$\pm$  0.04 \\
IAC80    & 2014/02/06 &  0.82 &  10.3 & -0.61$\pm$  0.07 \\ 
IAC80    & 2014/03/01 &  0.79 &  21.0 & -0.97$\pm$  0.03 \\
IAC80    & 2014/03/06 &  0.80 &   6.6 & -0.91$\pm$  0.08 \\
IAC80    & 2014/03/08 &  1.00 &  27.1 & -1.34$\pm$  0.05 \\
IAC80    & 2014/03/16 &  1.23 &  34.6 & -1.29$\pm$  0.05 \\
IAC80    & 2014/03/17 &  0.81 &   8.0 & -0.65$\pm$  0.05 \\
TNT      & 2014/03/25 &  1.53 &  29.9 & -1.05$\pm$  0.01 \\
IAC80    & 2014/03/26 &  0.86 &   8.2 & -1.22$\pm$  0.05 \\
TNT      & 2014/03/28 &  1.32 &  23.8 & -1.83$\pm$  0.01 \\
IAC80    & 2014/03/29 &  1.35 &  15.0 & -1.14$\pm$  0.05 \\
INT      & 2014/04/08 &  1.24 &   9.3 & -0.72$\pm$  0.05 \\
IAC80    & 2014/04/24 &  1.11 &   9.9 & -1.06$\pm$  0.07 \\
\hline       
\end{tabular}
\end{center}
\label{PLindex}
\end{table}    

\section{OPTICAL POWER DENSITY SPECTRUM}
\label{pdspec}

We compute the power density spectrum (PDS) of the de-reddened and de-trended
light curves (see Section\,\ref{lcurves}; with the mean de-reddened flux added
to preserve the flux values) using the Lomb-Scargle method to compute the
periodograms \citep{Press92} and  the same normalization method as is commonly
used in X-ray astronomy, where the power is normalized to the fractional RMS
amplitude per  Hertz \citep{Klis89}. To compute the PDS we use the constraints
imposed by the Nyquist frequency, the typical duration of each observation and
use the recipe given in  \citet{Horne86} to calculate the number of independent
frequencies. We then bin and fit the PDS in logarithmic space
\citep{Papadakis93} where the errors in each bin are determined from the
standard deviation of the points within each bin.  The white noise level is
subtracted by fitting the highest frequencies ($>$300\,Hz) with a white noise
plus red noise model.  As an example of the typical PDS, in Fig.\,\ref{pds}a we
show the PDS of the  light curve taken on March 8 and in Table\,\ref{PLindex} we
give the power-law index obtained for each light curve. We find that the PDS of
the flare light curves is dominated by a red-noise component  with a  median
power-law index of --1.2, typical of aperiodic activity in X-ray binaries and
X-ray transients in outburst and quiescence, which have a power-law index in the
range -1.0 to -2.0  \citep{Zurita03, Shahbaz03, Hynes03a, Hynes03b, Shahbaz04,
Shahbaz05, Shahbaz10, Shahbaz13}. 

In order to determine the significance of possible peaks above the red-noise
level, we use a Monte Carlo simulation similar to \citet{Shahbaz05}. We generate
light curves with exactly the same sampling and integration times as the real
data. We start with a model noise light curve generated using the method of
\citet{Timmer95},  with a power-law index as determined from the PDS of the
observed data and then add Gaussian noise using the errors derived from the
photometry.  We compute  1000 simulated light curves and then calculated the 99
per cent confidence  level at each frequency taking into account a realistic
number of independent trials \citep{Vaughan2005}.  As an example, in
Fig.\,\ref{pds}a  we show the 99 per cent confidence level contour, which  rules
out any significant peaks in the PDS.

We also compute the PDS of the multi-band ULTRACAM data taken on April 29, which
shows strong active/passive state behaviour. We find the power-law index of the
PDS to be --1.43$\pm$0.05, --1.34$\pm$0.07 and --1.10$\pm$0.05 in the 
\sloanr, \sloang and \sloanu bands respectively, which implies that there is
more high frequency variability at shorter wavelengths.

\begin{figure*}
\centering
\includegraphics[angle=0,width=14.cm]{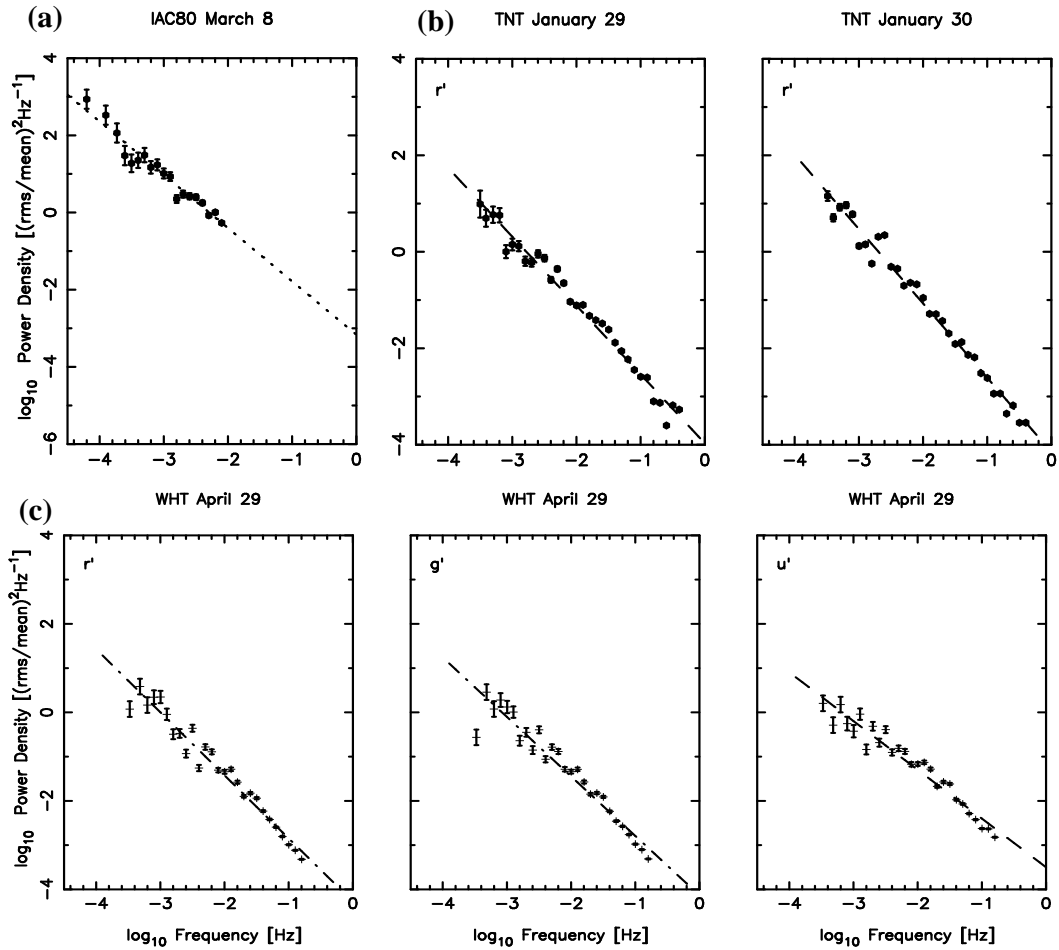}
\caption{
(a) An example of the typical power density spectrum (PDS) of the IAC80 and INT
light curves. The PDS shown is for the \sloanr-band light curve taken on March 8
and the solid line represents the 99 per cent confidence level, which  rules out
any significant peaks in the PDS.
(b) The PDS of the \sloanr-band TNT light
curves taken on January 29 and 30.  (c) The PDS of the \sloanr\ and \sloang\ and
\sloanu-band WHT light curves taken on
April 29. In each plot the  dot-dashed lines show a power-law fit to the PDS.}
\label{pds}
\end{figure*}

For the light curves that show mode-switching behaviour (January 29, 30 and
April 29), we compute the PDS of light curves during the active  and passive
states to see if they are different. 
We use the flare light curves computed in Section\,\ref{lcurves} with
the mean de-reddened flux added to preserve the flux values. In order to isolate
the passsive- and active states, we determine the  cut-off using the histogram
of the \sloanr-band flux values, where the cut-off is the flux  value that
divides the bimodal distribution (see Fig.\,\ref{subsec}). For the WHT data the
times in the \sloanr-band light curve corresponding to  flux values greater than
or less than  the cut-off value allows us to define the time values in the
\sloang and \sloanu-band  active  and passive state light curves.
Although we find a number of interesting differences in the slope of the
active and passive state PDS, a Monte Carlo analysis shows
that the differences are due to the window function of the light
curves. 
We perform a simulation in which we simulated a red-noise 
light curve with the same time values, power-law index, mean and rms as 
the \sloanr\ band light curve.  We then selected active and passive sections 
(the same as what we observe) and the compute the power spectrum and 
determine the power-law index. These were then compared to the 
power-law index of the simulated \sloanr band PDS.  
One would expect the PDS of the active 
and passive light curves to be the same as the PDS of the whole light curve. 
However, we find that the power-law index of the active 
and passive light curves are very similar, but different to the 
power-law index of the whole light curve.

\section{PROPERTIES OF THE LARGE FLARE}
\label{flares}

The multi-colour data  taken with ULTRACAM allow us to determine the colour of
the large flare event and dips (see Fig.\,\ref{detrend}d). The March 16 light
curve starts off with relative low variability (active state) which develops
into a large flare event lasting at least $\sim$ 30\,min. In contrast the April
29 light curve shows  dips in flux. To determine the colour of the flare event
we use the flare light curves computed in Section\,\ref{lcurves}, where the 
secondary star's sinusoidal modulation is removed  and the mean de-reddened flux
value is added to preserve the flux values in each band. For the March 16 data
we calculate the average active state and near-peak   flux values for the 
\sloanr, \sloang and \sloanu-band light curves. The flux of the flare is then
obtained by subtracting the active state flux from the near-peak flux value. 
For the April 29 data we calculate the average active  and passive state flux
values. The flare of the dips is then  obtained by subtracting the active state
flux from the passive state flux  (see Table\,\ref{fluxes}).

In an attempt to interpret the broad-band spectral properties of the flare
event, we compare the de-reddened colours with the prediction for  different
emission mechanisms, such as a blackbody, an optically-thin layer of hydrogen
and synchrotron emission. The most likely model for a thermal flare is emission
from an optically thin layer of recombining hydrogen, which is  the mechanism
generally accepted for stellar flares in single stars and Cataclysmic
Variables \citep[see][and references within]{Kerr14}. 
We therefore determine the continuum
emission spectrum of an LTE slab of hydrogen for different baryon densities,
$N_{\rm H}$ and temperatures $T_{\rm H}$. 
For each model we then compute the given emission
spectrum and the expected flux density ratios $f_{u'}/f_{g'}$ and
$f_{g'}/f_{r'}$ using the synthetic photometry package \textsc{synphot} 
(\textsc{iraf/stsdas}).  Given the intrinsic model flux we can then determine
the corresponding radius ($R_{\rm H}$) of the region that produces the observed
de-reddened flux at a given  distance of 1.368\,kpc \citep{Deller12}. 

In Fig.\,\ref{colcol} we show the colours of the March 16 active  and
flare-state data and the expected colours for different emission models. 
We can see that the blackbody or synchrotron emission models do not 
reproduce the observed flux ratio's and so these models can be ruled out.
For the
March 16 light curve, the active state flux ratios (see Table\,\ref{fluxes})
correspond to a region with $N_{\rm H}\sim 10^{25}$\,nucleons\,cm$^{-2}$ and
$T_{\rm H}\sim 6600$\,K. The flux ratio of the flare (near-peak flux minus
active state flux) corresponds to a region with $N_{\rm H}\sim
10^{23}$\,nucleons\,cm$^{-2}$, $T_{\rm H}\sim 8400$\,K and  $R_{\rm H}\sim
0.35$\Rsun.  Hence we find that during the large  flare event there is a
decrease in baryon density and an increase in temperature, suggesting that the
flare emission becomes hotter and more optically thin. Using the 
distance of 1.368\,kpc \citep{Deller12} and the corresponding filter bandwidth,
the optical peak luminosity of the large flare is 4.9, 2.9 and
3.3$\times10^{32}$\ergs\ in the \sloanu, \sloang\ and \sloanr-bands
respectively. Although we do not observe the full decay of the flare event, we
estimate that the  flare lasts $\sim$\,1000\,s  ($\sim$\,0.06 in phase) with an
energy of  6$\times10^{35}$\,ergs in the combined  \sloanu, \sloang\ and
\sloanr-bands.

For the April 29 light curve, the active state flux ratios  correspond to a
region with $N_{\rm H}\sim 10^{25}$\,nucleons\,cm$^{-2}$ and $T_{\rm H}\sim
6900$\,K (similar to the active state in the  March 16 light curve). The flux
ratio of the dips (i,e. the flux ratio between bands after subtracting the
active state flux from the passive state flux in each band) corresponds to a
region with $N_{\rm H}\sim 10^{30}$\,nucleons\,cm$^{-2}$, $T_{\rm H}\sim
4600$\,K and $R_{\rm H}\sim 0.35$\Rsun.   Note that if we assume no reddening 
the colour ratios  change by at most 6 per cent.

\begin{table*}
\caption{Mean flux values of the WHT light curves taken on March 16 and April
29. For the March 16 event the mean active, near peak fluxes are given. The
flare flux is calculated by subtracting the near-peak flare flux from the mean
active state flux and we also give the optical luminosity of the flare. For the
April 29 light curves the mean active  and passive state fluxes are given as
well as the  X-ray (0.5--10\,keV) to optical (\sloanu, \sloang\ and \sloanr)
luminosity ratio, where the X-ray active  and passive state luminosities are
2.85$\times 10^{33}$\ergs\ and  5.70$\times 10^{32}$\ergs\ respectively
\citep{Linares14c}
}
\begin{center}
\begin{tabular}{llcccc}\hline 
\textsc{ut}
Date      &   Band           & Active  state &  Near-peak    & Peak$-$Active & $L_{\rm OPT}$ \\
          &               & flux  (mJy)    &   flux (mJy)    & flux (mJy)         & ($\rm erg\,s^{-1}$)  \\ \hline
March 16   & $f_{u'}$        & 0.515(2)   & 1.369(7   & 0.855(9)    & 4.9$\times 10^{32}$ \\
           & $f_{g'}$        & 0.398(1)   & 0.868(4)  & 0.471(4)    & 2.9$\times 10^{32}$ \\
	   & $f_{r'}$        & 0.973(2)   & 1.921(8)  & 0.948(8)    & 3.3$\times 10^{32}$ \\   
	   & $f_{u'}/f_{g'}$ & 1.294(7)   & 1.576(12) & 1.825(15)   &  \\          
	   & $f_{g'}/f_{r'}$ & 0.409(1)   & 0.452(3)  & 0.497(2)    &  \\ \\ 
           &                 & Active  state & Passive  state & Active$-$Passive  &  $L_{\rm X}/L_{\rm OPT}$ \\
           &                 & flux  (mJy)   &   flux (mJy)   & flux (mJy)        &   Active\, \hspace{1mm} \,Passive\\ \hline
April 29   & $f_{u'}$        & 0.563(2)      & 0.538(1)       & 0.025(2)          & 9.0\, \hspace{3mm} \,1.9 \\
	       & $f_{g'}$        & 0.720(1)      & 0.632(1)       & 0.087(2)          & 6.5\, \hspace{3mm} \,1.5  \\
	       & $f_{r'}$        & 1.375(2)      & 1.174(1)       & 0.202(3)          & 6.2\, \hspace{3mm} \,1.4  \\
           & $f_{u'}/f_{g'}$ & 0.523(1)      & 0.539(1)       & 0.283(21)         &  \\
           & $f_{g'}/f_{r'}$ & 0.782(2)      & 0.851(1)       & 0.432(8)          &  \\
\hline     \hline
\end{tabular}
\end{center}
\label{fluxes}
\end{table*}    

\begin{figure}
\centering
\includegraphics[angle=-90,width=8.cm]{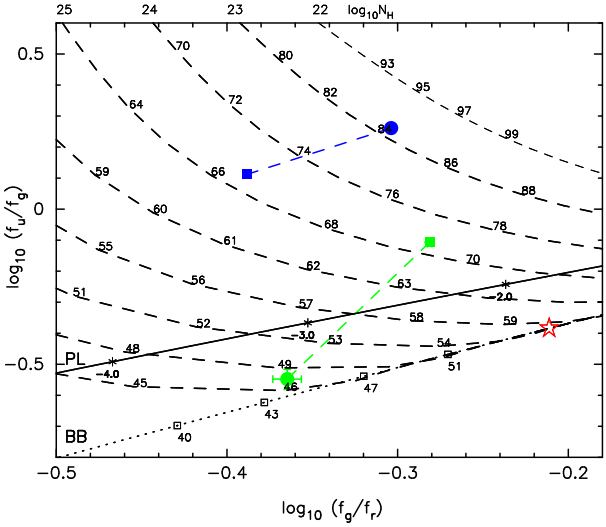}
\caption{The colour-colour diagram of the WHT light curves of \target\ taken  on the March 16 (blue) and April 29 (green). The colour during the active state is represented by a solid square, whereas the colour of the large flare event and dips is represented by a solid  circle. The  dashed lines show optically-thin hydrogen slab models  for  different baryon column densities $\log N_{\rm H}$, where the numbers mark the temperature in 100\,K units.  From right to left we show models with  $\log N_{\rm H}=22-30$\,nucleons\,cm$^{-2}$. The solid line shows a power-law  model ($F_{\nu} \propto \nu^{\alpha}$) with indices $\alpha$=--2.0 to --4.0 marked as stars. The dotted line shows a blackbody model marked in 
100\,K units as open squares. The large red star shows the colour of a G5 main sequence star.
}
\label{colcol}
\end{figure}
						     
\section{DISCUSSION}

\subsection{The flaring activity}

\citet{Patruno14}  reported the X-ray and UV/optical signatures of a state
change in \target\ during mid 2013, which coincided with the disappearance of
radio pulsations and the X-rays show a softer spectrum than during the
millisecond pulsar state, indicating a change in the origin of the X-ray
emission. $U$-band variability of $\sim$\,0.3 mag on time-scales of $\sim$\,1\,h
was seen as well as X-ray flux changes by up to a factor of $\sim$100 on
time-scales less than $<$10\,s.  During the previous active phase in 2002, slow
(hours) and rapid (minutes to seconds) optical flickering was observed
\citep{Bond02}. Our light curves observed in 2014 during the accretion-powered
phase show  rapid flaring events on time-scales of minutes with amplitudes of
$\sim$0.1--0.5\,mag and/or large flare events  on time-scales of $\sim$5--60 min
with amplitudes of $\sim$0.5--1.0\,mag (see Section\,\ref{flaring}). The PDS of
the flare light curves has a median power-law index of --1.2 (see
Section\,\ref{pdspec}), typical of X-ray binaries \citep{Zurita03, Shahbaz03,
Shahbaz04, Shahbaz05, Shahbaz10, Shahbaz13}. 

Since the accretion disc  is sandwiched by a hot, optically thin, fully ionized
corona, two possible origins for the flares are (1) the optically thick,
geometrically thin accretion disc or (2) from  the corona, leading to enhanced
energy release over a short period of time.  In the context of the propeller
effect, the inner edge of the accretion disc rotates slower than the neutron
star magnetosphere and in-falling matter is no longer able to accrete onto the
neutron star, but instead is blown out of the system by a propeller
\citep{Illarionov75}, leading to flaring activity.  This has been used to
explain the large luminosity variations in the outburst tails of
SAX\,J1808.4-3658 \citep{Campana08}. Indeed the sporadic large flaring events
observed in \target\ during its accretion-powered phase could also be explained
in terms of a propeller effect, leading to the ejection of hot optically thin 
fireballs, as is observed in some cataclysmic variables \citep{Pearson03}. 

For the large flare event observed on March 16 (lasting at least $\sim 30$\,min)
where we have multi-colour information, we find that  the event arises from a
relatively optically thin region with $\sim 8700$\,K, $N_{\rm H}\sim
6\times10^{22}$\,nucleons\,cm$^{-2}$ and a radius of  $\sim 0.35$\Rsun; the
out-of-flare state is cooler and more optically thick (see Fig.\,\ref{colcol}).
The energy of the flare in the optical bandpass of 6$\times10^{35}$\,ergs (see
Section\,\ref{flares}) is two orders of magnitude  larger than the optical
luminosity of flares produced by chromospheric activity in flare stars
\citep{Walkowicz11}. \citet{Church04} have measured the radial extent of the
accretion disc corona (ADC) in low mass X-ray binaries. They find  the ADC
extends from 20,000\,km in the faintest sources to 700,000\,km in the
brightest,  typically 15 per cent  of the accretion disc radius. In this
context, our data suggest that the large flares arise from a large fraction of
the accretion disc, most likely from the accretion disc corona.

\subsection{The X-ray/UV correlations}

Power-law correlations between the optical and X-ray luminosities in an X-ray
binary are naturally expected. The optical emission in low-mass X-ray binaries
is generally thought to arise in the outer accretion disc as the result of X-ray
reprocessing and the luminosities are correlated in the form $L_{\rm \nu; V}
\propto L_{\rm X}^{\beta}$ ($\rm V$ refers to the $V$-band)
\citep{vanParadijs94}. The index $\beta$ depends on the surface brightness of
the accretion which is given by $S_\nu \propto T^{\alpha}$, where
$\beta=\alpha/4$ \citep{vanParadijs94}. In the $V$-band $\alpha\sim2.0$ and so
$\beta\sim0.5$ \citep{vanParadijs94}. It can be shown that $\alpha$ increases
with decreasing wavelengths;   in the UV-band, $\alpha\sim3.7$ ($\beta
\sim0.9$),  in the $V$-band $\alpha\sim2.7$ ($\beta \sim0.7$)  and  in the   
$K$-band  $\alpha\sim1.2$ ($\beta\sim0.3$).  \citet{Russell06}  show that
similar correlations are  expected when the optical flux originates in the
viscously heated disc because both the X-ray and optical fluxes are linked
through the mass accretion rate.  For a  viscously-heated steady-state disc,
with outer disc temperatures of 8000--12,000\,K, $\beta$ is dependent on
wavelength and  increases at shorter wavelength; $\beta\sim0.3$ in the $K$-band,
$\beta\sim0.5$ in $V$-band and $\beta\sim0.60$ in the $U$-band. Note that the
power-law index due to an irradiated accretion disc is steeper than that due to
a viscously heated disc, for a given waveband. \citet{Russell06} have quantified
the disc and jet contribution from the optical/infrared and X-ray (2--10 keV)
correlation in X-ray binaries in the hard state. From an observational point of
view a strong optical/X-ray luminosity correlation from quasi-simultaneous data
has been observed in neutron-star X-ray binaries in the hard state, which takes
the form  $L_{\rm \nu; OIR} \propto L_{\rm X}^{0.63\pm0.04}$ and holds over
seven orders of magnitude in $L_{\rm X}$ (where $\rm OIR$ covers the $B$ to
$K$-band). The authors concluded that X-ray reprocessing dominates the
optical/near-infrared emission in neutron-star X-ray binaries  in the hard
state, with possible contributions from a jet (only at high luminosity) and a
viscously heated disc.

From Fig.\,\ref{lxlopt} we find the UV and X-ray luminosities are strongly
correlated with a power-law index $\sim 1.0$ in the  UV bands (see
Table\,\ref{index:lxlopt}). This is steeper than what is expected from a
viscously heated disc (0.6) and is more in line with the index expected  from an
irradiated accretion disc (0.9).

\subsection{The rectangular dips}

X-ray mode-switching behaviour observed as fast transitions between 
passive and active states (rectangular dips), has been detected  in \target\
\citep{Tendulkar14, Archibald14} and other similar systems \citep{Linares14c}.
These  dips, which are rectangular, are distinct from the dipping activity
observed in some  high inclination low-mass X-ray binaries, called X-ray
dippers, where intense dipping activity at specific orbital phases are observed
due to  azimuthal  absorbing material located close to the accretion disc bulge
\citep{White95}.  \textit{XMM-Newton} observations of PSR\,J1824--2452I, with an
X-ray luminosity $10^{35}$\ergs,  showed rectangular dips with ingress and
egress times of $\sim$ 200\,s and dip durations  of up to a few thousand seconds
\citep{Ferrigno14}. \textit{Chandra} observations in quiescence showed
mode-switching  behaviour with states lasting for $\sim$10\,h and transition
times of less than 500\,s \citep{Linares14b}. Dips which show a spectral
hardening, ingress and egress times of $<$10\,s and  durations between 200 and
800\,s have also been observed  in \textit{XMM-Newton} observations of XSS
J1227.0--4859 \citep{deMartino13}.    \citet{Linares14c} has shown that all
three redbacks observed so far which have transitioned to the accretion phase
(\target, XSS\,J1227.0--4859 and PSR\,J182--2452I)  show X-ray mode-switching,
with little or no change in X-ray photon index.  For XSS J1227.0--4859 and
M28-I, where there are measurements of $N_{\rm H}$ during the active  and
passive states,  no significant change in $N_{\rm H}$ is observed
\citep{deMartino13,Linares14a}.   \textit{NuSTAR} observations of \target\ in
October 2013, with an average X-ray luminosity of $5.8\times10^{33}$ \ergs, show
sharp-edged, rectangular, flat-bottomed dips  with durations of 30 to 1000\,s,
ingress and egress times of 30 to 60\,s and typical separations of $\sim$ 400\,s
\citep{Tendulkar14}.  The X-ray luminosity of the active  and passive states
only changes by a factor of $\sim $5 \citep{Linares14c}. 

Our high time-resolution optical observations show sharp-edged, rectangular dips
(see Section\,\ref{dips}), which is the first time that the optical analogue of
the X-ray active/passive state mode-switching  has been observed. The histogram
of the  optical  mid-dip position as a function of orbital phase is relatively
flat (see Fig.\ref{subsec}), suggesting that the dips are  uniformly and
randomly distributed with orbital phase, similar to what has been observed in
the X-rays. The timescale properties (ingress/egress, separation etc) of the
optical and X-ray dips are very similar, which suggests that we are seeing
echoes of the X-ray variability. 
We observe an
increase in the $f_{u'}/f_{g'}$ and $f_{u'}/f_{g'}$ colour ratio during the
dips, which  is  due to the disappearance of a red spectral component (see
Fig.\ref{zoom}).  The X-ray photon index of the active  and passive state X-ray
data does not change significantly \citep{Linares14c}.  
The decrease in the X-ray luminosity between the 
active and passive states is a factor of $\sim $5 
\citep{Linares14c}, much larger than the 
decrease in optical luminosity between the active and passive states; 
$\sim$10 per cent;
the decrease in the X-ray-to-optical luminosity ratio 
during the active and passive state of a factor between 4.4 and 4.7 
(see Table\,\ref{fluxes}).

\citet{Bogdanov14}  recently presented the results of a multi-wavelength study of
\target\ during it's low-mass X-ray binary state. B-band light curves taken with
a cadence  of 13\,s during late 2013 and 2014 clearly show the secondary
star's heated ellipsoidal modulation with superimposed flaring activity, similar
to our optical light curves (see Section\,\ref{lcurves}). The authors also
obtained considerable simultaneous \textit{XMM-Newton}  X-ray and B-band light curves
taken with a time resolution of 10\,s. They find
that none of the B-band light curves show the mode switching behaviour as seen
in the X-rays. At first light this seems to contradict our findings, where we
clearly observe optical mode switching in some of our high-time resolution light
curves. However, it should be noted that in the ULTRACAM data, where mode
switching is observed, the mode switching is strongest in the \sloanr band and is
significantly weaker in the \sloang and \sloanu-bands; the mode switching flux changes by
a factor of 17, 14 and 5 per cent in the \sloanr, \sloang and \sloanu-band light curves
respectively (see Fig.\,\ref{lcurve}d and Table\,\ref{fluxes}).  Therefore, it
is not surprising that the B-band light curves obtained by \citet{Bogdanov14} do
not show clear discernible mode switching  behaviour.

\citet{Bogdanov14} extensive \textit{XMM-Newton} observations allowed them 
to and observe the source in various states: $\sim$20, $\sim$70 and $\sim$10 
percent  of the time in the low- (passive), high- (active) and flare-state
respectively.  Our ground-based optical observations total about 96 hr.  During
this time, we find the source to be in the flare-state for  $\sim$\,10 hr.
Although we identify passive/active state behaviour in only  6\,hr of data, it
should be noted that the poor time-resolution of the  IAC80 and INT data does
not allow us to resolve the rapid flaring. It is  clear that more extensive
high-time resolution optical data is needed before we  can compare the
properties of the optical and X-ray state changes.

\subsubsection{Proposed scenario for the optical and X-ray mode-switching}

The details of the interaction between the rotating magnetosphere of the neutron
star and the inner accretion flow are  complicated (e.g., see
\citealt{Uzdensky04, Lai14}. However, it is generally assumed that accretion
onto the neutron star's surface occurs when a geometrically thin disc (
truncated at the `magnetospheric radius' \rma; where the ram pressure of the
gas balances the pressure of the magnetic  field, \citealt{Pringle72}) is less
than the corotation radius, \rco\  (the radius where the disc Keplerian frequency
matches the neutron star spin frequency, which is 24\,km for \target).  If the 
inner edge of the disc extends inside the pulsar's light cylinder radius 
(\rlc; where the field lines
attached to the neutron star synchronously rotate with the star at a
velocity equal to the speed of light) but  lies outside  \rco,  ` propeller accretion' can occur
\citep{Illarionov75}. In this case the pressure of infalling material is
balanced by the magnetic field of the neutron star outside  \rlc\ (\rma$>>$\rco)
and the infalling matter does not reach the neutron star surface.  The resulting
centrifugal barrier created by the rotating  neutron star  expels the material
from the system and  the radio pulsar's emission is quenched  \citep{Spruit93,
Rappaport04}.  However, it is worth noting that the radiation pressure of a
radio pulsar does not  necessarily have to disrupt the disc and the disc can
remain stable outside \rlc\ \citep{Eksi05}. 

The recent detection of X-ray pulsations in \target\ during the accretion
active state \citep{Archibald14, Bogdanov14}  implies that channeled accretion,
similar to that seen in higher-luminosity accreting millisecond X-ray pulsars,
is occurring at a much lower accretion rate, implying  that the inner edge of
the accretion disc lies close to \rco\ and a propeller does not form. In this
case material can accumulate near \rco\ and non-stationary accretion can occur
as  matter piles up around the intrinsically unstable magnetospheric boundary. 
Accretion discs accreting onto the magnetosphere of a rotating star can end up
in a trapped state, in which the inner edge of the disc stays near \rco. The
captured material can form a quasi-spherical shell \citep{Pringle72, Shakura12}
or  a new disc structure, known as a `dead-disc' and episodic accretion can
occur \citep{D'Angelo10, D'Angelo12}.   As noted by \citet{Patruno14},
variations in the  mass-accretion rate of the accretion disc can move the inner
disc radius by a factor of two or three. The viscous time-scale defines the
time-scale for this drift and to reach a viscous time-scale of 10--100\,s 
requires a region with an annulus of radius $\sim$10--100\,km. 
The thermal time-scale is given by $t_{\rm therm} \sim (H/R)^2\ t_{\rm vis}$ 
(where $H$ and $R$ are the height and radius of the disk) and for an  inner
thin  disk with $H/R<$0.02  implies $t_{\rm therm} <$few seconds, much lower
than the time-scales observed.
Thus, in
principle, fluctuations in the mass accretion rate can move \rma\ outside
\rlc\,,which allows the radio pulsar to turns on, triggering a transition to the
passive state, where the lower X-ray luminosity is produced by a shock between
the pulsar wind and innermost accretion flow.  An increase in the mass accretion
rate pushes the inner edge of the disc back inside the light cylinder and turns
off the radio pulsar. The X-ray pulsations are only observed in the active state
and not in the passive state light curves,  which suggests that the switching
between the passive  and active states results in transitions between a
non-accreting pure propeller mode and an accreting trapped-disc
mode\citep{Archibald14}.

To explain the optical/X-ray timing and spectral properties of the passive  and
active state light curves, we propose a scenario in which the optical flux
during the passive  and active states arise from the viscously-heated and X-ray
reprocessed disc,  respectively, as well as from the irradiated secondary star.
High energy photons (presumably X-rays)  produced in the inner parts of the
accretion disc, very close to the compact object,  photoionize and heat the
surrounding outer regions of gas in the accretion disc and secondary star. These
regions later recombine and cool, producing lower energy photons. This results
in reprocessed optical and UV radiation  arising from a volume of significant
spatial extent,  that is imprinted with the same variability as the X-ray
signal.  
The steeper PDS at shorter wavelengths suggests more
high frequency variability arising from reagions closer to the compact object,
as expected.
An increase in the mass accretion rate pushes the inner edge of the
disc  inwards, past some critical radius which lies most likely
near \rlc, and the radio pulsar turns off. During this accretion-powered
active state, matter is accreted onto the neutron star surface via the magnetic
poles, resulting in X-ray pulsations \citep{Archibald14},  as well as
irradiating the outer parts of the accretion disc. The buildup and release of
mass in a trapped disc at some critical radius  leads to  clumpy accretion onto
the neutron star's surface, as depicted  by the rectangular flat-bottom light
curves. A decrease in the mass accretion rate moves the inner edge of the
accretion disc outwards allowing  the radio pulsar to turn on, triggering a
transition to the rotation-powered passive state, where the lower X-ray
luminosity is produced by a shock between the pulsar wind and  innermost
accretion flow and the  viscously-heated  accretion disc.  


 \section{CONCLUSIONS}

Below we list the main results of this paper.

\begin{enumerate}

\item
Our  long-term  time-resolved \sloanr-band  light curves taken between 2014
January and April show an underlying sinusoidal modulation due to the irradiated
secondary star and accretion disc. We also observe superimposed rapid flaring on
time-scales less than minutes with amplitudes of $\sim$0.1--0.5\,mag, and
further additional large flare events on time-scales of $\sim$5--60 min with
amplitudes of $\sim$0.5--1.0\,mag. The power density spectrum of the flare light
curves is dominated by a red-noise component with a power-law index typical of
aperiodic activity in X-ray binaries.

\item The UV and X-ray luminosities are strongly correlated suggesting that the
same emission mechanism is powering part of the X-ray and UV emission. The
power-law index of the X-ray/UV luminosity correlation is consistent with an
irradiated accretion disc.  

\item 
For the large flare event observed on  March 16 where we have colour
information, we find that  the event arises from a relatively optically thin
region with temperature $\sim 8700$\,K, $N_{\rm H}\sim
6\times10^{22}$\,nucleons\,cm$^{-2}$ and a radius of  $\sim 0.35$\Rsun, possibly
from the accretion disc corona.

\item
On some nights we also observe sharp-edged, rectangular, flat-bottomed dips 
randomly distributed with orbital phase. These rectangular dips are similar to
the mode-switching behaviour between active  and passive  luminosity states observed
in the X-ray light curves of other similar millisecond pulsars. The dips have a
median duration and ingress/egress time of  $\sim 250$\,s and $\sim 20$\,s
respectively and we observe an increase in the blue colour ratio during the
dips. This is the first time that the optical analogue of the X-ray
mode-switching has been observed.   

\end{enumerate}

To explain the timing and spectral properties of the active  and passive state
light curves, we propose a scenario in which  the optical flux during the
passive  and active states arise from the viscously-heated and X-ray
reprocessed disc and changes in the inner disc radius, which lies near \rco.  An increase in the
mass accretion rate leads to an accretion-powered active state, where matter is
accreted onto the neutron star surface via magnetic poles, resulting in X-ray
pulsations \citep{Archibald14}, as well as irradiating the outer parts of the
accretion disc.   A decrease in the mass accretion rate moves the inner disc's
edge outwards, triggering a transition to the rotation-powered passive state,
where the lower X-ray luminosity is produced by a shock between the pulsar wind
and the innermost accretion flow and the  viscously-heated  accretion disc. 
The optical/X-ray timing  properties of the passive  and active
state light curves, can be explained in  terms of transitions between a
non-accreting pure propeller mode accretion  and an accreting trapped-disc mode accretion
with X-ray reprocessing, during which clumpy accretion from a trapped inner
accretion disc near the corotation radius results in  rectangular, flat-bottomed optical and X-ray light
curves.

\section*{ACKNOWLEDGEMENTS}

TS would like to thank Mallory Roberts and Rene Breton for useful discussions. 
This research has been supported by the Spanish Ministry of Economy and
Competitiveness (MINECO) under the grants AYA2010-18080, AYA2012--38700 and
AYA2013-42627. PRG is supported by a Ram\'on y Cajal fellowship
RYC--2010--05762. VSD, SPL and ULTRACAM are supported by STFC grant
ST/J001589/1. Based on scheduled observations made with the William Herschel
Telescope and the Isaac Newton Telescope operated on the island of La Palma by
the Isaac Newton Group in the Spanish Observatorio del Roque de Los Muchachos of
the Instituto de Astrof\'{i}sica de  Canarias and on observations made with the
IAC80 telescope operated on the island of Tenerife by the Instituto de
Astrof\'{i}sica de Canarias in the Spanish Observatorio del Teide.

\footnotesize{

}

\end{document}